  \documentclass[pra,showpacs,twocolumn]{revtex4}

  \usepackage{amsmath}
  \usepackage{amssymb}
  \usepackage{bm}
  \usepackage{hyperref}
\usepackage{natbib}

  \usepackage{graphicx}
  \def\dfrac{\displaystyle\frac}

 \newcommand {\e}{{\rm e}}
\renewcommand {\i}{{\rm i}}
\renewcommand {\Re}{\mathop{\mathrm{Re}}\nolimits}
\renewcommand {\Im}{\mathop{\mathrm{Im}}\nolimits}
\newcommand {\rot}{\mathop{\mathrm{rot}}\nolimits}

\renewcommand {\phi}{\varphi}
\newcommand {\erf}{\mathop{\mathrm{erf}}\nolimits}

\newcommand{\eL}{\varepsilon_\parallel}
\newcommand{\eT}{\varepsilon_\perp}
\newcommand{\eps}{\varepsilon}

\begin{document}
  \title{Spontaneous radiation of a finite-size dipole emitter in hyperbolic media}

  \author{Alexander N. Poddubny,$^{1,2}$ Pavel A. Belov,${}^{1,3}$ and Yuri S. Kivshar,${}^{1,4}$}
    \affiliation{$^{1}$Department of Photonics and Optoinformatics, National University of Information Technology, Mechanics and Optics,
St.~Petersburg 197101, Russia\\
  $^{2}$Ioffe Physical-Technical Institute of the Russian Academy of Science,  St.~Petersburg 194021, Russia\\
$^{3}$School of Electronic Engineering and Computer Science,
Queen Mary University of London, Mile End Road, London E1 4NS, UK\\
  $^{4}$Nonlinear Physics Center and Center for Ultrahigh-bandwidth Devices for Optical Systems (CUDOS),
  Research School of Physics and Engineering, Australian National University, Canberra ACT 0200, Australia}


\begin{abstract}
We study the radiative decay rate and Purcell effect for a finite-size dipole emitter placed in a homogeneous uniaxial medium.
We demonstrate that the radiative rate is strongly enhanced when the signs of the longitudinal and transverse dielectric constants
of the medium are opposite, and the isofrequency contour has a hyperbolic shape. We reveal that the Purcell enhancement factor
remains finite even in the absence of losses, and it depends on the emitter size.
\end{abstract}

\maketitle

\section{Introduction}

Purcell effect is the enhancement of the spontaneous emission for a source placed in the resonant cavity as compared to that in vacuum~\cite{Purcell}.  The engineering of the radiative lifetime is now extensively studied in a variety of different 
systems including  metallic particles~\cite{tanaka2010,shubina2010,Meinzer2010,glazov2011arXiv},
 microcavities~\cite{khitrova2006,Dousse2010,Yang2011}, and  metamaterials~\cite{Xie2009,hughes2009,narimanov2010,narimanov2010b,Noginov2010}.

The huge Purcell factor for a point dipole embedded in the so-called hyperbolic medium has been reported in Ref.~\onlinecite{narimanov2009}.
This system, namely an uniaxial medium where the transverse $\varepsilon_{xx}=\varepsilon_{yy}\equiv\eT$ and longitudinal
$\eps_{zz}=\eL$ dielectric constants have the opposite signs, is characterized with the hyperbolic isofrequency contours in the wavevector space~\cite{lindell2001,smith2003,Felsen}, see also the insets in Fig.~\ref{fig:1}.  Wave propagation and refraction in the hyperbolic medium reveals its unusual optical properties, as compared to the uniaxial medium with the ellisoidal isofrequency surface~\cite{lindell2001,smith2004,Smith2004b,Cai2008}.
 The radiative rate of the point dipole in such a medium diverges, and it remains finite only due to the inevitable losses~\cite{narimanov2010,narimanov2010b,Noginov2010}.

In this work, we consider a finite-size light source, such as a quantum dot, placed in a homogeneous uniaxial medium
with $\eL'\eT'<0$. We demonstrate that for the spatially distributed source the radiative rate does not diverge even for 
vanishing losses, but it depends strongly on the source size instead. The maximum enhancement of the Purcell factor 
can be roughly estimated as $(\lambda/a)^3$, where $\lambda$ is the light wavelength is vacuum, and $a$ is the 
characteristic size of the source.

 Importantly, the hyperbolic medium employed in our calculations is not only a hypothetic theoretical model. It appears as
  an effective medium in the homogenization of the artificial photonic structures--- metamaterials, the layered structures 
  consisting of alternating dielectric and metallic slabs~\cite{Belov2006,Salandrino2006,Jacob2006}, as well as 
the structures crated by a mesh of metallic wires~\cite{elser2006,Yao2008,narimanov2009b,nefedov2011arXiv}.
In these metamaterials, one also have to take into account the strong spatial dispersion of the effective dielectric constants~\cite{belov2003,elser2007b} and the excitation of plasmons~\cite{elser2007c,elser2008,orlov2011arXiv}.

The paper is organized as follows. In Sec.~\ref{sec:model} we introduce our model and outline the calculation technique. 
The results of calculation are summarized in Sec.~\ref{sec:results}, whereas Sec.~\ref{sec:concl} concludes the paper. 
Details of some calculations are given in Appendix~\ref{sec:A}.

\section{Model}
\label{sec:model}

We consider a spherical light source (e.g. a quantum dot) embedded into an anisotropic homogeneous medium
characterized by the dielectric tensor $\hat\varepsilon$.
Equation for the electric field reads
\begin{equation}\label{eq:E}
 \rot\rot \bm E=q_0^2\bm D\:,
\end{equation}
where $q_0=\omega/c$, $\omega$ is the wave frequency, and $c$ is the speed of light in vacuum.
The displacement vector includes the background contribution and resonant polarization of the emitter
$\bm P$:
\begin{equation}
 \bm D=\hat\varepsilon \bm E+4\pi \bm P\:,
\end{equation}
and the nonzero components of the dielectric tensor are
\begin{equation}
 \varepsilon_{xx}=\varepsilon_{yy}=\eT,\quad \varepsilon_{zz}=\eL\:.
\end{equation}
We write the phenomenological material equation for the polarization as~\cite{Pilozzi2004,Ivchenko2005}
\begin{equation}\label{eq:P}
  \bm P=\frac{d^2\Phi(r)}{\hbar(\omega_0-\omega)}\int d^3 r'\Phi(r')\bm E(\bm r')\:.
\end{equation}
Here $\omega_0$ is the resonance frequency, $d$ is the effective matrix element of the dipole moment of the emitter, and
the function $\Phi(r)$ characterizes the spatial distribution of the emitter polarization. In what follows, we use
$\Phi(r)$ in the simple Gaussian form
\begin{equation}
 \Phi(r)=\frac{\sqrt{2}\exp(-r^2/2a^2)}{4\pi^{3/2}a^3}\:,
\end{equation}
so that $\int d^3r\Phi(r)=1$. Equation~\eqref{eq:P} is similar to the material relation for 
the excitonic polarization of the semiconductor quantum dot, see Ref.~\onlinecite{Ivchenko2005}.

The radiative lifetime $\tau$ is related to the complex eigenfrequency $\omega$ of
the homogenous system Eq.~\eqref{eq:E}--Eq.~\eqref{eq:P} as~\cite{Ivchenko2005,Khitrova2002,Goupalov2003}
\begin{equation}\label{eq:res3}
 \frac{1}{\tau}=-2\Im \omega\:.
\end{equation}
To find $\tau$ we apply the Fourier transform,
\begin{equation}
 \bm E(\bm r)=\int\frac{d^3k}{(2\pi)^3}\bm E_{\bm k}\e^{\i\bm k\bm r}\:,
\end{equation}
and obtain
\begin{equation}\label{eq:e2}
 [q_0^2\hat\varepsilon\bm E_{\bm k}-k^2\bm E_{\bm k}+\bm k(\bm k\cdot \bm E_{\bm k})]=-\frac{4\pi q_0^2d^2\Phi_k }{\hbar(\omega_0-\omega)}\int\frac{d^3k'}{(2\pi)^3}\bm E_{k'}\Phi_{k'}\:,
\end{equation}
where
\begin{equation}\label{eq:Phi_k}
 \Phi_k=\int d^3r\e^{-\i\bm k\bm r}\Phi(r)=\e^{-k^2a^2/2}\:.
\end{equation}
In the derivation of Eq.~\eqref{eq:e2}, we took into account that
the function $\Phi_{\bm k}$ depends only on the absolute value of the vector $\bm k$.
 Equation~\eqref{eq:e2} can be rewritten as
\begin{equation}\label{eq:e3}
 \bm E_{\bm k}=-\frac{4\pi  q_0^2d^2}{\omega_0-\omega}\Phi_k\hat G_{\bm k}\bm \Lambda\:,
\end{equation}
where we introduced the Green function in the $\bm k$-space
\begin{equation}\label{eq:G}
 \hat G_{\bm k}=(\hat M_{\bm k})^{-1},\quad M_{\bm k,\alpha\beta}=q_0^2\varepsilon_{\alpha\beta}-k^2\delta_{\alpha\beta}+k_\alpha k_\beta
\end{equation}
and defined a new variable, $\bm \Lambda=\int \bm E_{\bm k}\Phi_kd^3k/(2\pi)^3$.
Multiplying both the parts of Eq.~\eqref{eq:e2} by $\Phi_k$ and integrating over $\bm k$, we
 obtain the matrix equation for the complex eigenfrequencies $\omega$,
\begin{equation}\label{eq:resL}
 (\omega-\omega_0)\bm \Lambda=\hat R\bm \Lambda,\quad \hat R=\frac{4\pi q_0^2d^2}{\hbar} \int\frac{d^3k}{(2\pi)^3}\Phi_k^2\hat G_{\bm k}\:.
\end{equation}
We note that the matrix $\hat R$ in the right-hand side of Eq.~\eqref{eq:resL} generally depends on the frequency $\omega$. However, we are interested in the weak coupling regime, when the interaction of the emitter with light can be treated as a perturbation~\cite{kavbamalas}, and we set $\hat R(\omega)=\hat R(\omega_0)$ in Eq.~\eqref{eq:resL}. Taking into account  that the matrix $\hat R$ is diagonal due to
the symmetry of the problem, we obtain the spontaneous emission times
\begin{equation}\label{eq:tau}
 \frac{1}{\tau_\alpha}=-\frac{8\pi d^2q_0^2}{\hbar }\int\frac{d^3k}{(2\pi)^3}
\Im G_{\alpha,\bm k}\Phi_k^2,\quad \alpha=x,y,z\:.
\end{equation}
The times $\tau_x=\tau_y$ and $\tau_z$ describe the decay of the source, initially polarized in the plane $xy$ and along the $z$ axis, respectively.
To find the decay rates, one should substitute the explicit expressions for the Green function,
\begin{align}\label{eq:Gk}
 G_{\bm k,zz}&=\frac{1}{\eL}\frac{1-k_z^2/(q_0^2\eT)}{q_0^2-k_{\perp}^2/\eL-k_{\parallel}^2/\eT}\:,\\\nonumber
G_{\bm k,xx}&=\frac{1}{k_{\perp}^2\eT}\left\{\frac{k_y^2}{q_{0}^2-k^2/\eT}+
\frac{k_x^2[1-k_\perp^2/(q_0^2\eL)]}{q_0^2-k_{\perp}^2/\eL-k_{\parallel}^2/\eT}\right\}\:,
\end{align}
into Eq.~\eqref{eq:tau}. As follows from Eq.~\eqref{eq:Gk}, the axial dipole couples both with TE (ordinary) and TM (extraordinary) waves, while
the orthogonal dipole couples only with TM  waves. The mode dispersion can be determined from the poles of the Green functions and is illustrated by insets of Fig.~\ref{fig:1}(a) and (b).

The convergence of the integrals \eqref{eq:tau} is assured by the
rapidly decaying function $\Phi_k^2$. Thus, the cutoff is naturally provided by the source size, $k_{\rm max}\sim 1/a$, similarly as it happens in the  nonrelativistic theory of the Lamb shift~\cite{feinberg1974}. We note, that the realistic metamaterial such as the wire medium is characterized by the lattice constant $a_0$. If the emitter size  is smaller than the spacing between the wires, $a<a_0$, our approach is not applicable, and the cutoff is provided at $k_{\rm max}\sim 1/a_0$. It was shown in Ref.~\onlinecite{maslovski2011}, the enhancement of the density of TEM modes in the wire medium as compared to the TM modes in vacuum, is of the order of $1/(q_0a_0)^2$. This value provides the estimation of the Purcell factor for the wire medium.

\section{Results and discussions}
\label{sec:results}

The integral in Eq.~\eqref{eq:tau} can be readily calculated numerically. In case of the source size smaller than its wavelength, one can also obtain explicit analytical expressions (see Appendix~\ref{sec:A} for more details),

\begin{widetext}

\begin{align}\label{eq:main}
 \frac{1}{\tau_z}=\frac{q_0^3d^2}{\hbar}\Bigl\{\Im
\frac{\arctan\sqrt{\eps-1}-\sqrt{\eps-1}}{\sqrt{\pi}\eT
(\eps-1)^{3/2}
 (q_0a)^3}+\Im\frac{(\eps-2)\arctan\sqrt{\eps-1}+\sqrt{\eps-1}}{\sqrt{\pi}
(\eps-1)^{3/2}
 q_0a}+\frac{4}{3}\Re\sqrt{\eT}
\Bigr\}\:,\\
\frac{1}{\tau_x}=\frac{1}{\tau_y}=
\frac{q_0^3d^2}{\hbar}\Bigl\{\Im
\frac{\sqrt{\eps-1}-\eps\arctan\sqrt{\eps-1}}{2\sqrt{\pi}\eT
(\eps-1)^{3/2}
 (q_0a)^3}+\Im\frac{\eps^2\arctan\sqrt{\eps-1}-\eps\sqrt{\eps-1}}{2\sqrt{\pi}
(\eps-1)^{3/2}
 q_0a}+\Re\frac{\eL+3\eT}{3\sqrt{\eT}}\nonumber
\Bigr\}\:,
\end{align}

\end{widetext}

where
\begin{equation*}
 \eps=\eL/\eT\:.
\end{equation*}
Eqs.~\eqref{eq:main} present the central result of this work. They are valid for an arbitrary complex values of
$\eL$ and $\eT$ provided that $q_0a\sqrt{|\eL|}\ll 1$,\quad $q_0a\sqrt{|\eT|}\ll 1$. The experimentally observed decay kinetics of the emitter will be determined by the excitation conditions. In case when the direction of the dipole moment is fixed and makes the angle $\theta$ with the symmetry axis $z$, the decay will be biexponential with the initial slope  given by
\begin{equation}
 \frac{1}{\tau(\theta)}=\frac{\cos^2\theta}{\tau_z}+\frac{\sin^2\theta}{\tau_x}\:.
\end{equation}
In the isotropic medium, where $\eT=\eL=\kappa$ all the rates \eqref{eq:main} reduce to
\begin{equation}\label{eq:iso}
\frac{1}{\tau}=\frac{4q_0^3d^2}{3\hbar}\Re\sqrt{\kappa}-\frac{d^2}{3\hbar\sqrt{\pi}a^3}\Re\frac{1}{\kappa}\:.
\end{equation}
In the transparent medium ($\kappa''=0$) the first term in Eq.~\eqref{eq:iso} reduces to the textbook result for the spontaneous emission rate~\cite{Delerue2004}. The second term describes the energy losses due to the heating of the medium~\cite{barnett1996,tomas1999} similarly as for a dipole placed in the pore in metal~\cite{glazov2011arXiv}.
This term controls the decay rate, when the real part of the dielectric constant is negative.

\begin{figure}[t]
 \includegraphics[width=\linewidth]{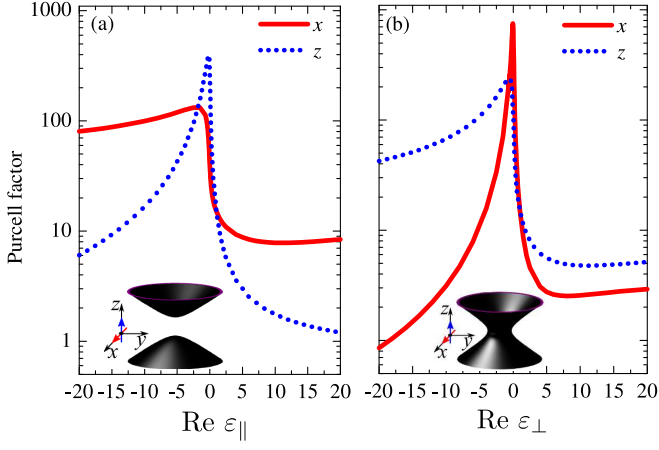}
\caption{(Color online) Purcell factor relative to vacuum as a function of (a) $\eL'$ for
 $\eT'=1$ and (b) $\eT'$ for  $\eL'=1$.
 Solid and dashed lines correspond to the dipole oriented along the $x$ and $z$ axes,
  respectively. The insets show schematically the isofrequency surfaces in
 the $\bm k$-space for $\eT'\eL'<0$.
Calculation was performed for $q_0a=0.1$ and $\eL''=\eT''=0.1$.
}\label{fig:1}
\end{figure}

In the anisotropic medium with  vanishing losses, i.e.
$
 \eL''=\eT''\to 0\:
$
Eqs.~\eqref{eq:main} reduce to

\begin{equation}\label{eq:main_X}
\frac{1}{\tau_\alpha}=
\begin{cases}
                              W_{\alpha}^{(0)}, &\eT'>0,\eL'>0\:,\\
                              W_{\alpha}^{(0)}+
\dfrac{W_{\alpha}^{(1)}}{q_0a}+
 \dfrac{W_{\alpha}^{(3)}}{(q_0a)^3}
, &\eT'>0,\eL'<0\:,\\
-\dfrac{W_{\alpha}^{(1)}}{q_0a}+
 \dfrac{W_{\alpha}^{(3)}}{(q_0a)^3}	
, &\eT'<0,\eL'>0\:,\\
0, &\eT'<0,\eL'<0\:,\\
                              \end{cases}
\end{equation}
where  $\alpha=x,y,z$,
\begin{align}\label{eq:main_Z}
 W_{z}^{(0)}&=\dfrac{4q_0^3d^2\sqrt{\eT'}}{3\hbar},
W_{x}^{(0)}=\frac{4q_0^3d^2(\eL'+3\eT')}{3\hbar\sqrt{\eT'}},\\
\nonumber
 W_{z}^{(1)}&=-\dfrac{\sqrt{\pi|\eT'|}q_0^3d^2(2|\eT'|+|\eL'|)}{2\hbar(|\eL'|+|\eT'|)^{3/2}},\nonumber\\
W_{x}^{(1)}&=\dfrac{\sqrt{\pi}q_0^3d^2|\eL'|^2}{4\hbar\sqrt{\eT'}(|\eL'|+|\eT'|)^{3/2}},\nonumber\\
 W_{z}^{(3)}&=\dfrac{\sqrt{\pi |\eT'|}q_0^3d^2}{2\hbar(|\eL'|+|\eT'|)^{3/2}
},\nonumber
W_{x}^{(3)}=\dfrac{\sqrt{\pi}q_0^3d^2|\eL'|}{4\hbar\sqrt{|\eT'|}(|\eL'|+|\eT'|)^{3/2}}
\end{align}
and $W_y\equiv W_x$.

Eq.~\eqref{eq:main_X} clearly demonstrates
nonanalytical behavior with $\eL$ and $\eT$.
When both constants are positive, we are dealing with
traditional uniaxial dielectric and transition rates do not depend on the dipole size.
For $\eL<0$  and $\eT<0$ all radiative rates are zero since the waves in such medium are evanescent and do not carry energy away from the source.
The most interesting regime is realized when longitudinal and transverse dielectric constants are of the opposite sign, $\eL\eT<0$.
In this case the emission rates are governed by the terms
$\propto 1/a^3$ and $\propto 1/a$ in \eqref{eq:main_X}, i.e. are determined by local field effects.
Counterintuitively, in the regime when $\eL\eT<0$ the local field is extended in the whole space\cite{Felsen} and thus controls the radiative emission. The detailed analysis of this peculiar field pattern will be presented elsewhere.
\begin{figure}[t]
 \includegraphics[width=\linewidth]{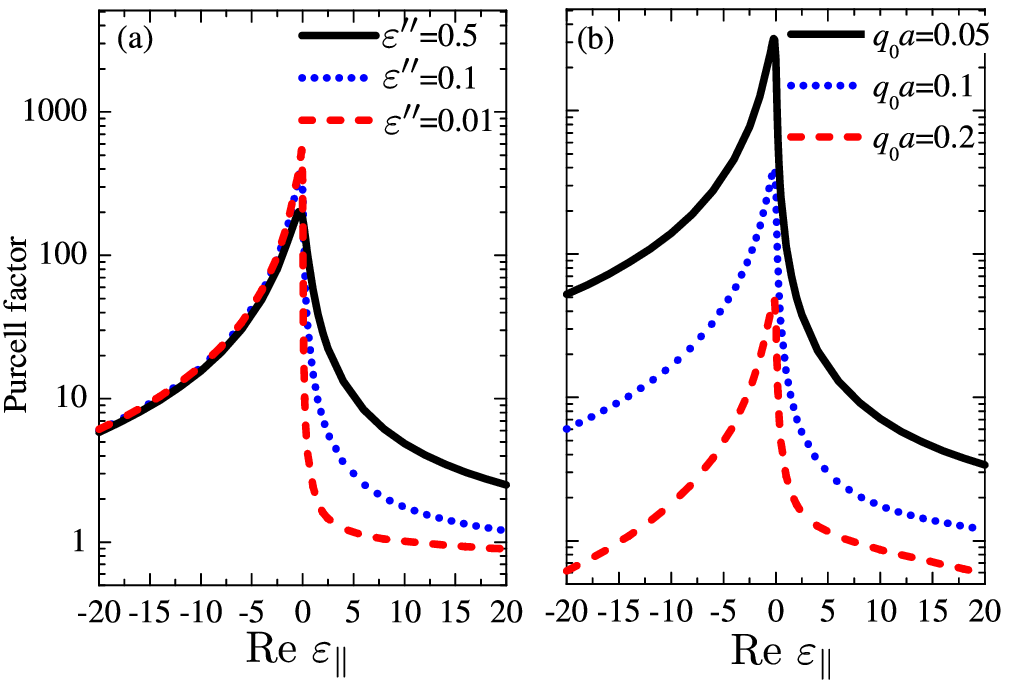}
\caption{(Color online) (a) Purcell factor as a function of $\eL'$ for
 $\eT''=\eL''=0.01$ (dashed curve),
$\eT''=\eL''=0.1$ (dotted curve),
$\eT''=\eL''=0.5$ (solid curve) and $q_0a$=1.
  (b)  Same as in (a) but for
 $q_0a=0.2$ (dashed curve),
$q_0a=0.1$ (dotted curve),
$q_0a=0.05$ (solid curve) and $\eT''=\eL''=0.1$.
The dipole is oriented along the $z$ axis.
}\label{fig:2}
\end{figure}


\begin{figure}[t]
 \includegraphics[width=7.0 cm]{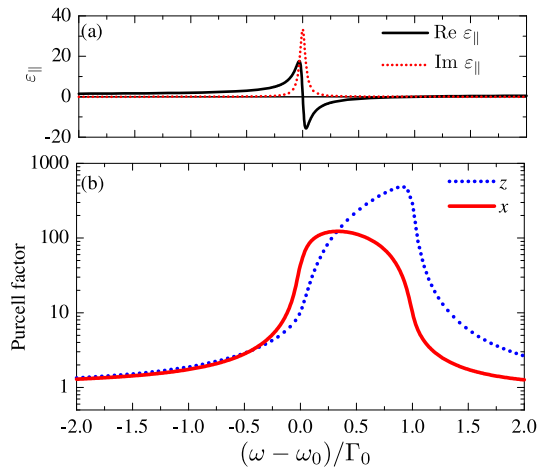}
\caption{(Color online)
Frequency dependence of the Purcell factor in the medium with
$\eT=1$ and $\eL$ given by Eq.~\eqref{eq:res}.
(a) Functions $\eL'(\omega)$ (solid curve) and
$\eL''(\omega)$ (dotted curve).
(b) Purcell factor for the dipoles oriented along $x$ (solid curve)
and along $z$ (dotted curve). Calculation was performed for $q_0a=0.1$, $\eL^{(0)}=1$, $\Gamma/\Gamma_0=0.03$.
}\label{fig:3}
\end{figure}

The results of numerical calculation of the transition rates based on Eqs.~\eqref{eq:tau} and
\eqref{eq:Gk} are summarized in Figs.~\ref{fig:1} to \ref{fig:3}. The rates are normalized to their 
values at $\eT=\eL=1$, which yields the Purcell factor with respect to vacuum.
Figure~\ref{fig:1} shows the Purcell factor for the different dipole orientations as a function of 
(a) $\eL'$ and (b) $\eT$. In agreement with Eqs.~\eqref{eq:main}, the rates drastically increase when 
the real part of one of the dielectric  constants becomes negative. Interestingly, the largest enhancement 
in Fig.~\ref{fig:1}(a) is achieved when  $\eL'$ is negative but small, i.e. when $\eL'<0$ and $|\eL'|\ll 1$. 
In agreement with this result, the leading terms in Eqs.~\eqref{eq:main_X} for the transition rate are proportional 
to $\sqrt{\eT'}/(a^3|\eL'|^{3/2})$ for $|\eL'|\gg 1$. This  justifies additionally the fact that the observed enhancement 
is the local field effect, because for large values of $|\eL'|$ the local field is screened, and so the effect is suppressed. Similar analysis applies for the dependence of the transition rates on $\eT'$, see Fig.~\ref{fig:1}(b). We notice 
that the analytical results~\eqref{eq:main}  describe all the curves in Fig.~\ref{fig:1} with a precision better than $5\%$.

Figure~\ref{fig:2} shows how the Purcell factor dependence on $\eL$ changes (a) with losses
and (b) with source size $a$.
From Fig.~\ref{fig:2}a we conclude that the losses smear the nonanalytic behaviour of Purcell factor when $\eL'$ crosses zero and reduce maximum value of Purcell factor. On the other hand, in the regime when $\eL'>0$, $\eT'>0$, the losses lead to the growth of the decay rate. This is similar to the isotropic case, Eq.~\eqref{eq:iso} and is related to the heating of the medium by the emitted field.
 Fig.~\ref{fig:2}b shows, that the Purcell factor is very sensitive to the dipole size. It is quickly suppressed when the size increases, in agreement with $1/a^3$ and $1/a$ terms in Eqs.~\eqref{eq:main}.

Figure~\ref{fig:3} illustrates the frequency dependence of the Purcell factor in the medium where
\begin{equation}\label{eq:res}
 \eL(\omega)=\eL^{(0)}+\frac{\Gamma_0}{\omega_0-\omega-\i\Gamma}
\end{equation}
and $\eT=1$. Comparing Fig.~\ref{fig:3}(a) and Fig.~\ref{fig:3}(b),  we observe that the largest enhancement is achieved
in the spectral region where $\eL'(\omega)$ is negative, but small, which agrees with our analysis of Fig.~\ref{fig:1}. 
As a result, the positions of the maxima of the curves in Fig.~\ref{fig:3}(b) are blue-shifted from the resonance energy $\omega_0$.

\section{Conclusions}
\label{sec:concl}

We have developed the theory of the Purcell effect for spherical dipole emitters embedded in homogeneous uniaxial media, 
taking into account a finite size of the emitter and losses in the surrounding medium. We have obtained analytical 
expressions for the decay rates in the case when the emitter size is much smaller than the wavelength of radiation.  
We have revealed that, when the real parts of the longitudinal and transverse dielectric constants $\eL'$ and $\eT'$ 
are of the opposite sign (i.e. for the so-called hyperbolic media), the radiative decay rate depends strongly on the 
emitter size, and it diverges when the size vanishes. This enhancement is related to the peculiar
nature of the local field in such systems, which spatially extends to infinity. The largest Purcell factor is achieved when
$\eL'\eT'<0$ and  the absolute values of the  dielectric constants are much smaller than unity, since the screening of the 
local electric field in this case is minimal. Our theory has been developed for a homogeneous medium, but it can also 
provide a qualitative insight into the problem of the spontaneous emission in metamaterials.

 \acknowledgments

This work was supported by the Ministry of Education and Science of the Russian Federation, RFBR
and Dynasty Foundation (Russia), EPSRC (UK), and the Australian Research Council (Australia).
The authors acknowledge numerous illuminating discussions with I. Iorsh, E.L. Ivchenko, S.I. Maslovski, 
A.S. Potemkin, and C.R. Simovski.

\appendix

\section{Analytical expressions for the decay rates}\label{sec:A}

In this Appendix, we present the details of the derivation of Eq.~\eqref{eq:main}\:.

First, we substitute Eqs.~\eqref{eq:Gk} into Eq.~\eqref{eq:tau} and introduce the spherical coordinates $(k,\theta,\phi)$ in the $\bm k$-space.
Integration over the azimuthal angle $d\phi$ and over $dk$ can be performed analytically, and  yields

\begin{widetext}

\begin{multline}\label{eq:theta_gen}
 \frac{1}{\tau_{\rm rad}^{z}}=\frac{q_0^3d^2}{\hbar}\Im\int_0^\pi d\theta\Bigl\{
-\frac{\chi(\theta)\sin\theta\cos^2\theta}{2\sqrt{\pi}\eL\eT(q_0a)^3}+\frac{\chi^2(\theta)\sin^3\theta}{\sqrt{\pi}\eL^2q_0a}+\frac{\i \chi^{5/2}(\theta)\sin^3\theta}{\eL^2}\e^{-(q_0a)^2\chi(\theta)}
\bigl[1+{\rm erf}[\i q_0a \sqrt{\chi(\theta)}]\bigr]
\Bigr\}\:,
\end{multline}
\begin{multline}\label{eq:theta_genX}
 \frac{1}{\tau_{\rm rad}^{x}}=
\frac{1}{\tau_{\rm rad}^{y}}=
\frac{q_0^3d^2}{\hbar}\Im\int_0^\pi d\theta\Bigl\{\frac{\i\sin\theta\sqrt{\eT}}{2}\e^{-\eT (q_0a)^2}[1+{\rm erf}[\i q_0a \sqrt{\eT}]-
\frac{\chi(\theta)\sin^3\theta}{4\sqrt{\pi}\eL\eT(q_0a)^3}\\+\frac{\sin\theta
(\eL-\chi\sin^2\theta)}{2\sqrt{\pi}\eL\eT q_0a}+\frac{\i \chi^{3/2}(\theta)\sin\theta}{2\eL\eT}\e^{-(q_0a)^2\chi(\theta)}
\bigl[1+{\rm erf}[\i q_0a \sqrt{\chi(\theta)}]\bigr][\eL-\chi(\theta)\sin^2\theta]
\Bigr\}\:,
\end{multline}

\end{widetext}

where
\begin{equation}
 \frac1{\chi(\theta)}=\frac{\sin^2\theta}{\eT}+
\frac{\cos^2\theta}{\eL}\:,
\end{equation}
and the error function is defined as
\[
 \erf(x)=\frac{2}{\sqrt{\pi}}\int_0^x\e^{-t^2}\:.
\]

We consider the case when the source size is very small, so that
the condition
\begin{equation}\label{eq:smallloss}
  q_0a |\sqrt{\chi(\theta)}|\ll 1,\quad q_0a |\sqrt{\eT}|\ll 1,\quad q_0a |\sqrt{\eL}|\ll 1
\end{equation}
is satisfied for all values of $\theta$. If the real parts of the dielectric constants have the same sign, 
the conditions~\eqref{eq:smallloss} are easily satisfied for small $q_0 a$. However,  if $\eL'\eT'<0$
the quantity $\chi(\theta)$ may vanish. In this  case,  the first condition \eqref{eq:smallloss} will 
be still satisfied provided the imaginary parts of the dielectric constants are sufficiently high.
Under the conditions~\eqref{eq:smallloss} the exponential functions in Eqs.~\eqref{eq:theta_gen} can be replaced 
by unity, and the error functions can be neglected.

After that simplifications, the integration over $\theta$ can be performed analytically, and it gives Eqs.~\eqref{eq:main}. 
Our numerical analysis shows that Eqs.~\eqref{eq:main} hold even when $\eT''$, $\eL''$ vanish, provided that the last two conditions \eqref{eq:smallloss} remain valid. In this case, the terms  in Eqs.~\eqref{eq:theta_gen} and \eqref{eq:theta_genX} proportional to $\exp[-(q_0a)^2\chi(\theta)]$ are rapidly oscillating, so that their contribution to the integrals becomes small.

%

\end{document}